\documentclass[a4paper, twocolumn, nofootinbib, showpacs, amsmath, amssymb]{revtex4}

\usepackage{graphicx}

\begin{document}

\title{ Field propagation in a stochastic background space: \\The rate of light incoherence in stellar interferometry  }

\author{Michael~Maziashvili }
\email{maziashvili@gmail.com} \affiliation{Center for Elementary Particle Physics, ITP, Ilia State University, 3/5 Cholokashvili Ave., Tbilisi 0162, Georgia}

\begin{abstract}

We present Hilbert space representation for a relatively broad class of minimum-length deformed quantum mechanical models obtained by incorporating a space-time uncertainty relation into quantum mechanics. The correspondingly modified field theory is used for estimating the deviation of the light incoherence rate from distant astrophysical sources from the standard case.     

\end{abstract}




\maketitle

\section{Introduction}

In the framework of general relativity, gravitational field is described by the space-time metric \cite{LL}-\S 82. Thus, the measurement of the gravitational field is tantamount to the measurement of space-time distances. Taking into account the quantum-mechanical fluctuations, it was shown in a number of papers that there inevitably exists an intrinsic uncertainties in measuring the space-time distances \cite{Osborne:1949zz, Anderson, Wigner:1957ep, Salecker:1957be, Regge:1958wr, Peres:1960zz, Mead:1964zz, Mead:1966zz, Karolyhazy:1966zz}. Otherwise speaking, space-time metric undergoes inherent quantum-mechanical fluctuations. Simply, on the dimensional grounds one could get an idea that in the Minkowski background, the rate of fluctuation of a length scale $l$ should have the form $\delta l = \beta l_P^{\alpha}l^{1-\alpha}$, since $l_P$ is the only quantity with dimensions of length that one can construct from the quantities $c, G_N, \hbar$ \cite{Planck} (here $\beta$ is a numerical factor of order unity). For a relatively recent review we refer the reader to reference \cite{Garay:1994en}. (In what follows we will use natural system of units: $c=\hbar=1$). As to the parameter $\alpha$, one cannot be very strict in defining its proper value. One can just require the values of $\alpha$ to be such that $\delta l \ll l$ for $l \gg l_P$. In what follows we will be interested in length scales much greater than the Planck length. The purpose of this paper is 1) to develop a systematic way for incorporating the  relation $\delta l = \beta l_P^{\alpha}l^{1-\alpha}$ into field theory and 2) use the modified field theory for estimating the coherence rate for the light coming from distant astrophysical sources. The idea proposed in \cite{Lieu:2003ee} to estimate the phase fluctuation accumulated by the plane wave $\exp(i[\omega t - \mathbf{k}\mathbf{x}])$ ($\omega = k$) over the time $t$ as $t\delta \omega$, where $\delta \omega$ is calculated by means of the relations $ \omega = 2\pi/\lambda,\, \delta \lambda = \beta l_P^{\alpha}\lambda^{1-\alpha}$, is clearly very dubious. So is an alternate idea to assume that over the length scale $l$ the path fluctuation for an electromagnetic wave caused by the background metric fluctuations should be estimated irrespective to its wavelength as $\delta l = \beta l_P^{\alpha}l^{1-\alpha}$ thus implying the phase fluctuation of the order of $\delta l /\lambda$ \cite{Ng:2004xr}. What one can say definitely on the bases of the above discussion is that the fluctuation in wavelength should be estimated as $\delta \lambda = \beta l_P^{\alpha}\lambda^{1-\alpha}$; however the question of how this fluctuation adds up over a length scale $l$ requires some theoretical framework (one may naturally expect the wavelength to have some bearing on this question).

We will abandon the above-mentioned "imaginative" concepts and will try to incorporate the relation $\delta l = \beta l_P^{\alpha}l^{1-\alpha}$ into quantum mechanics and hence into the field theory.

\section{Incorporating the relation $\delta l = \beta l_P^{\alpha}l^{1-\alpha}$ into the quantum mechanics}

We start off with a Minkowskian background the spatial distance of which undergoes the fluctuations quantified by the relation $\delta l = \beta l_P^{\alpha}l^{1-\alpha}$. The question of the possible space-time structure at the Planck scale is beyond the scope of our discussion, so that from the outset we assume that $l \gg l_P$. As a concomitant of these fluctuations the position uncertainty in Heisenberg uncertainty relation gets increased as 

\begin{equation}\label{mlunceryrel}
\delta X \,\geq\, \frac{1}{2\delta P} \,+\, \beta l_P^{\alpha}\delta X^{1-\alpha} ~. 
\end{equation} This modification can be understood as an immediate result of fluctuations (uncertainty) $\delta \lambda = \beta l_P^{\alpha}\lambda^{1-\alpha}$ in the de Broglie-wavelength of the incident particle by means of which we are measuring the position of the observed particle. So long as $\delta X \gg l_P$ we can rewrite Eq.\eqref{mlunceryrel} in the form

\begin{equation}\label{gur} \delta X \delta P  \,\geq\, \frac{1}{2} \,+\, \beta l_P^{\alpha}\delta P^{\alpha} ~. \end{equation}

\subsection{Hilbert space representation of Eq.\eqref{gur}}
\label{Hilbert}

To find a concrete Hilbert space representation of $\widehat{X},\,\widehat{P}$ operators, we start off with the deformed $\mathcal{QM}$

\begin{equation}\label{ertganz} \left[ \widehat{X},\, \widehat{P}  \right]  \,=\, i \left(1 \, +\, \beta \, l_P^\alpha \widehat{P}\,^\alpha \right) ~,  \end{equation} that is dictated by the form of Eq.\eqref{gur} (numerical factors of order unity are absorbed in $\beta$). This sort of QM is characterized with minimum position uncertainty of the order of \cite{Maslowski:2012aj} 

\[ \delta X \, \simeq \,  \left[\int\limits_0^{\infty} \frac{dP}{1 \, +\, \beta \, l_P^\alpha P\,^\alpha} \right]^{-1} \,=\, \frac{\beta^{1/\alpha}l_P}{\int\limits_0^{\infty} \frac{dq}{1 \, +\, q\,^\alpha} }~.  \] So, in the case $\alpha < 1$ the position uncertainty can reach zero while for $\alpha > 1$ it exhibits a nonzero minimum uncertainty in position that is seen immediately from the Eq.\eqref{gur} as well.

 A multidimensional generalization of Eq.\eqref{ertganz} can be written in the form  

\begin{eqnarray}\label{multdimcr}&& \left[\widehat{X}_i,\, \widehat{X}_j \right] \,=\, 0~,~ \left[\widehat{P}_i,\, \widehat{P}_j \right] \,=\, 0 ~, \nonumber \\&& 
 \left[\widehat{X}_i,\, \widehat{P}_j \right] \,=\, i\left\{  \Xi\left(\widehat{P}^2\right) \delta_{ij} \,+\, \Theta \left(\widehat{P}^2\right)\widehat{P}_i\widehat{P}_j \right\}~,~ 
\end{eqnarray} the Hilbert space representation of which can be constructed in terms of the standard $\widehat{\mathbf{x}},\, \widehat{\mathbf{p}}$ operators as

\begin{eqnarray}\label{hspacerep} \widehat{X}_i \, = \, \widehat{x}_i ~,  ~~ \widehat{P}_j \,=\,  \widehat{p}_j\xi\left(\widehat{\mathbf{p}}^2\right) ~. \end{eqnarray}
  
\noindent Let us work in the eigen-representation of operator $\widehat{\mathbf{p}}$: $\widehat{x}_i = i\partial/\partial p^i,\, \widehat{p}_j =p_j$. The simplest {\tt ansatz} would be to take

\begin{equation} \Theta \,=\, \frac{2\beta \,l_P^\alpha}{  \widehat{P}^{2-\alpha}}  ~,\nonumber \end{equation} thus from Eqs.(\ref{multdimcr}, \ref{hspacerep}) we get  

\begin{eqnarray}&& \left( \frac{\partial}{\partial p^i} \, p_j\xi\left(\mathbf{p}^2\right) \,-\,  p_j\xi\left(\mathbf{p}^2\right) \frac{\partial}{\partial p^i}  \right)\psi(\mathbf{p}) \,=\,   \nonumber \\&& \left( \delta_{ij}\xi\left(\mathbf{p}^2\right) \,+\,2p_ip_j \,\frac{d\xi\left(\mathbf{p}^2\right)}{d\mathbf{p}^2} \right)\psi(\mathbf{p})  \,=\, \nonumber \\&&  \left( \Xi\left(p^2\xi^2\right)\delta_{ij} +2\beta \, l_P^\alpha \,\frac{p_ip_j}{p^{2-\alpha}} \,\xi^\alpha \right)\psi(\mathbf{p})~, \nonumber  \end{eqnarray}

\noindent that is, 

\begin{eqnarray}&& \frac{d\xi\left(p^2\right)}{d p^2}  \,=\,   \beta \, l_P^\alpha  \,\frac{\xi^\alpha}{p^{2-\alpha}}~,~~ \Rightarrow  \nonumber \\&& \xi\left(p^2\right) \,=\, \left(1 \,-\,  \frac{2\beta(\alpha-1)}{\alpha} \, l_P^\alpha p^\alpha  \right)^{\frac{1}{1-\alpha}} ~.  \end{eqnarray} So we get

\begin{eqnarray}\label{pdefqmrep}  \widehat{X}_i \, = \, \widehat{x}_i ~, ~~~  \widehat{P}_j \,=\,  \widehat{p}_j \left(1 \,-\,  \frac{2\beta(\alpha-1)}{\alpha} \, l_P^\alpha \widehat{p}^\alpha  \right)^{\frac{1}{1-\alpha}} ~, \end{eqnarray} or in the eigen-representation of operator $\widehat{\mathbf{p}}$

\begin{eqnarray}\label{pdefqmrepeigenp}  \widehat{X}_j \, = \, i \, \frac{\partial}{\partial p_j} ~, ~~~ \widehat{P}_j \,=\,  p_j\left(1 \,-\,  \frac{2\beta(\alpha-1)}{\alpha} \, l_P^\alpha p^\alpha  \right)^{\frac{1}{1-\alpha}} ~, \end{eqnarray} with scalar product containing a cut-off on $p$ when $\alpha > 1$ 

\begin{eqnarray}\label{scalarproduct} \langle \psi_1 | \psi_2 \rangle = \int\limits_{p^{\alpha} \,<\, \alpha/2\beta(\alpha-1)l_P^\alpha}d^3p \, \,\psi^*_1(\mathbf{p})\psi_2(\mathbf{p})~,  \nonumber\end{eqnarray} and without cut-off on $p$ when $\alpha < 1$

\begin{eqnarray} \langle \psi_1 | \psi_2 \rangle = \int d^3p \, \,\psi^*_1(\mathbf{p})\psi_2(\mathbf{p})~.  \nonumber\end{eqnarray} In the case $\alpha=2$ one recovers the well-known result, see \cite{Kempf:1996nk, Kempf:1996fz}. Let us notice that the above-mentioned cutoff $p^{\alpha} \,<\, \alpha/2\beta(\alpha-1)l_P^\alpha$ when $\alpha > 1$ arises merely from the fact that when small $p$ runs over this region - large $P$ covers the whole region from $0$ to $\infty$, see Eqs.(\ref{pdefqmrep}, \ref{pdefqmrepeigenp}). In what follows we will use the abbreviation $\mathsf{PL}\mathcal{QM}$ for the Planck-length deformed quantum mechanics, Eqs.(\ref{multdimcr},\,\ref{pdefqmrep}).

\section{Field theory in light of the $\mathsf{PL}\mathcal{QM}$}

Before proceeding let us notice that for in the stellar interferometry one usually deals with the natural light, there is no preferential polarization direction for the emitted field. So, we treat light signal as a scalar quantity, (that means to take account of the scalar potential only) and consider scalar field instead of the electromagnetic one.

Let us first consider $\mathsf{PL}\mathcal{QM}$ with $\alpha =2$. In this case we have \cite{Kempf:1996nk, Kempf:1996fz} 

\begin{equation}\label{deformedopmultid} X^i = x^i\,,~~~~P^i = \frac{p^i}{1-\beta l_P^2 \mathbf{p}^2}~. \nonumber \end{equation} Its Hilbert space realization in the $\mathbf{p}$ representation has the form

\begin{equation}\label{standardprepresentation} X^i\psi(\mathbf{p}) = i\partial_{p_i} \psi(\mathbf{p})\,,~ P^i\psi(\mathbf{p}) = \frac{p^i}{1-\beta l_P^2 \mathbf{p}^2}\,\psi(\mathbf{p})~, \nonumber \end{equation}  with the scalar product 

\begin{equation}\label{scalarproduct} \langle \psi_1 | \psi_2 \rangle = \int\limits_{\mathbf{p}^2 < 1/\beta l_P^2}d^3p \,\psi^*_1(\mathbf{p})\psi_2(\mathbf{p})~. \nonumber \end{equation} 

\noindent The modified field theory

\begin{eqnarray}\label{scaction}  \mathcal{A}[\varPhi] = - \int d^4x \, \frac{1}{2} \left[\varPhi\partial_t^2\varPhi  + \varPhi{\mathbf P}^2\varPhi \right] \,=\,  \nonumber \\ - \int d^4x \, \frac{1}{2} \left[\varPhi\partial_t^2\varPhi  + \varPhi\frac{-\Delta}{(1+\beta l_P^2\Delta)^2}\varPhi  \right] ~,\end{eqnarray}

\noindent results in the equation of motion of the form 

\begin{eqnarray}\label{eqofmot} 
\left(\partial_t^2 \,+\, \mathbf{P}^2  \right)\varPhi \,=\, \left(\partial_t^2 \,-\, \frac{\Delta}{(1+\beta l_P^2\Delta)^2}  \right)\varPhi \,=\,  \nonumber \\ \left( \partial_t^2  \,-\, \Delta \sum\limits_{n=0}^{\infty}\left(n + 1 \right)\left(-\beta l_P^2\Delta \right)^n  \right) \varPhi \,=\,0~.~ \end{eqnarray} Let us look at the spherical solutions $\varPhi(t,\, r)$. Recalling that the radial part of the Laplace operator in spherical coordinates has the form 

\[ \Delta \varPhi(t,\, r) \,=\, \frac{1}{r^2} \, \frac{\partial}{\partial r} \left( r^2\frac{\partial \varPhi}{\partial r} \right) ~, \] after making the substitution $\varPhi = \varphi /r $ one finds \[ \Delta \frac{\varphi}{r} = \frac{1}{r}\, \frac{\partial^2\varphi}{\partial r^2} ~,~\Delta^2  \frac{\varphi}{r} =  \frac{1}{r}\, \frac{\partial^4\varphi}{\partial r^4} ~, ~ \Delta^n  \frac{\varphi}{r} =  \frac{1}{r}\, \frac{\partial^{2n}\varphi}{\partial r^{2n}}~.  \] Considering a monochromatic wave 

\[  \varPhi(t,\, r)  \,=\, \frac{e^{-i(\omega t - k r)}}{r} ~, \] from Eq.\eqref{eqofmot} we find the modified dispersion relation

\begin{eqnarray}&& \omega^2 \,=\, \frac{k^2}{\left(1\,-\,\beta l_P^2k^2 \right)^2} ~,~ \Rightarrow  \nonumber \\&&    k^2 \,=\, \frac{\left(\sqrt{1 \,+\,4\beta l_P^2 \omega^2} \,-\,1 \right)^2}{4\beta^2 l_P^4\omega^2} ~. \label{moddisprel} \end{eqnarray}

Now let us look at $\mathsf{PL}\mathcal{QM}$ with $\alpha =1/2$. In this particular case from Eq.\eqref{pdefqmrep} one gets  

\begin{eqnarray}  \widehat{X}_i \, = \, \widehat{x}_i ~, ~~~  \widehat{P}_j \,=\,  \widehat{p}_j \left(1 \,+\,  2\beta l_P^{1/2} \widehat{p}^{1/2}  \right)^2 ~. \end{eqnarray} Hence the field equation of motion gets modified as

\begin{eqnarray}\label{modzraobisgant} &&
\left(\partial_t^2 \,+\, \mathbf{P}^2  \right)\varPhi \,=\, \nonumber\\&& \left(\partial_t^2 \,-\, \Delta \left[ 1+2\beta l_P^{1/2}(-\Delta)^{1/4}\right]^4  \right)\varPhi \,=\, 0~. \nonumber \end{eqnarray} Following the above discussion, now for the spherical, monochromatic-wave solution  

\[  \varPhi(t,\, r)  \,=\, \frac{e^{-i(\omega t - k r)}}{r} ~, \] we find 

\begin{eqnarray}\label{meoredisperstanafardoba}
\omega \,=\, k \left[ 1+2\beta l_P^{1/2}k^{1/2}\right]^2 ~,~~ \Rightarrow \nonumber \\ k \,=\, \frac{\left(1 \,-\, \sqrt{1+8\beta\sqrt{l_P\omega}} \right)^2}{16\beta^2l_P} ~.
\end{eqnarray}

\section{The degree of light coherence from distant celestial objects: van Cittert-Zernike formalism}

The light from the astrophysical source certainly cannot be strictly monochromatic for even the spectral lines for isolated atoms have a finite widths. In addition, the broadening of the spectral lines are caused because of motion of atoms (Doppler broadening) and also because of interaction/collisions between the atoms. In the case of real sources it is appropriate to talk about the wave-packet 

\begin{eqnarray}\label{wavepacket}  \varPhi(t,\, r) \,=\,  \int d\omega
\,a(\omega) \, \frac{e^{i\left[k(\omega)\,r-\omega
    t\right] }}{r} ~,  \end{eqnarray} where $k(\omega)$ is defined by Eq.\eqref{moddisprel} and the function $a(\omega)$ is understood to differ
appreciably from zero only within a narrow range around a mean frequency $\bar{\omega}$
\[\bar{\omega}-{\delta\omega\over 2} \leq\, \omega\, \leq
\bar{\omega}+{\delta\omega\over 2}~, ~~~~~{\delta\omega\over
\bar{\omega}}\ll 1~. \] If $\delta\omega$ is sufficiently small, the wave packet Eq.(\ref{wavepacket}) can be
interpreted as a plane wave with frequency $\bar{\omega}$, wave
number $k(\bar{\omega})$ and variable amplitude

\begin{eqnarray}\label{amplitude} A(t,\,r) = \int\limits_{\bar{\omega}-\delta\omega /2 }\limits^{
\bar{\omega}+\delta\omega / 2}d\omega \,
a(\omega)e^{i\left\{\left[k(\omega) -
      k(\bar{\omega})\right] \,r-\left[\omega - \bar{\omega}\right]
    t\right\}}~,\end{eqnarray}

\begin{eqnarray}\label{quasimonochromatic}  \varPhi(t,\, r) \,=\, A(t,\,r)\,\frac{e^{i\left[k(\bar{\omega})\,r-\bar{\omega} t\right] }}{r} ~.  \end{eqnarray} The width $\delta\omega$, determining the duration of the wave packet
$\delta t \simeq \delta\omega^{-1}$, is an important
characteristic for the interference effect. Namely, the interference effect
takes place when the path difference between the overlapping quasi-monochromatic
beams, Eq.\eqref{quasimonochromatic}, is less than the coherence length $\delta t$.

\begin{figure}[t]

\includegraphics[width=0.45\textwidth]{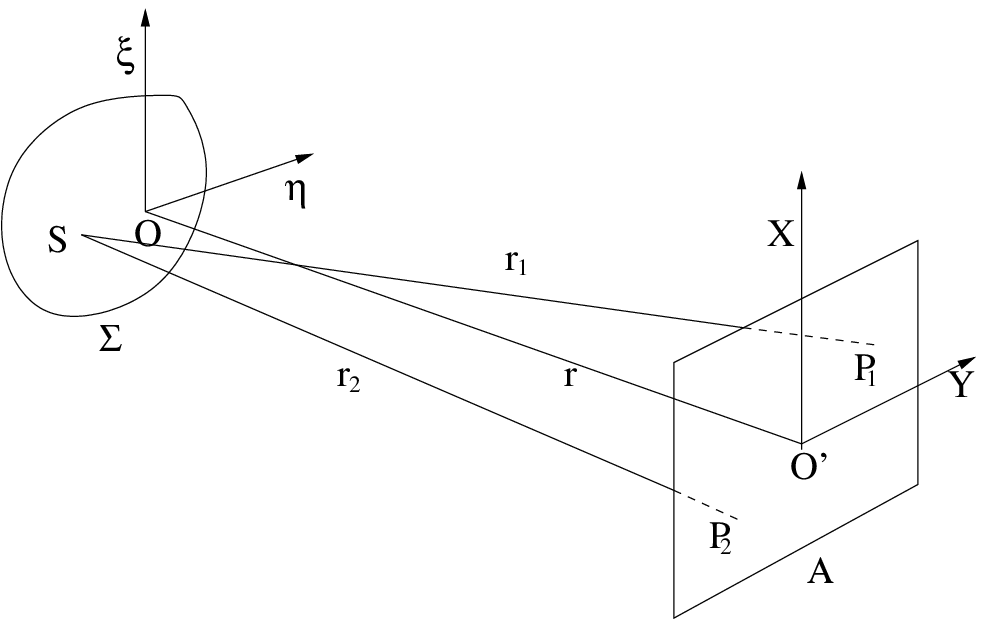}\\
\caption{ A screen $A$ illuminated by an extended
quasi-monochromatic incoherent source $\Sigma$. }
\end{figure}

Now we are in a position to follow van Cittert-Zernike theory to the mutual coherence of light from an extended quasi-monochromatic source \cite{BornWolf}. A screen $A$
is illuminated by an extended quasi-monochromatic incoherent
source $\Sigma$ taken for simplicity to be a portion of a plane
parallel to $A$, see Fig.1. The points $P_1$ and $P_2$ on the screen correspond to the interferometer slits. Before going on let us adopt the following natural assumptions. The linear dimensions of $\Sigma$ are small
compared to the distance $OO'$ between the source and the screen,
and the angles between $OO'$ and the line joining a typical source
point $S$ to $P_1$ and $P_2$ are small. Dividing the source into
elements $d\sigma_m$ centred on points $S_m$ and denoting
by $\varPhi_{m1}(t)$ and $\varPhi_{m2}(t)$ the disturbances at
$P_1$ and $P_2$ due to element $d\sigma_m$, for total disturbances
at these points one finds
\[\varPhi_{1,2}(t) = \sum\limits_m \varPhi_{m1,2}(t)~.\] Correlation
function between the light signals $\varPhi_{1}(t)$ and $\varPhi_{2}(t)$ takes the form (brackets $\langle ~\rangle$ denote time averaging)

\[\langle\varPhi_1(t)\varPhi^*_2(t) \rangle = \sum\limits_m
\langle\varPhi_{m1}(t)\varPhi^*_{m2}(t) \rangle +
\sum\limits_{m\neq n} \langle\varPhi_{m1}(t)\varPhi^*_{n2}(t)
\rangle~. \] The light signals
coming from different elements of the
source are  mutually incoherent, that is, there is
no correlation between $\varPhi_{m1}(t)$ and  $\varPhi_{n2}(t)$
when
$m\neq n$
\[\langle\varPhi_{m1}(t)\varPhi^*_{n2}(t) \rangle =
0~,~~~~\mbox{for}~~m\neq
n~.\] Using an explicit expression 
\begin{equation}\label{sphwaves}\varPhi_{m1,2}(t) =
A_m(t,\, r_{m1,2})\,\frac{e^{-i[\bar{\omega}t-k(\bar{\omega})r_{m1,2}]}}{r_{m1,2} }  ~, \end{equation} where $r_{m1,2}$ denote distances between the
elements $d\sigma_m$ and the points $P_{1,2}$, for the correlation function one finds  

\begin{eqnarray}\label{intcond} && \langle\varPhi_{m1}(t)\varPhi^*_{m2}(t)
\rangle   \,=  \nonumber \\&&  \left\langle A_m(t,\,r_{m1})A^*_m(t,\,r_{m2})
\right\rangle \,\frac{e^{ik(\bar{\omega})(r_{m1}-r_{m2})}}{r_{m1}\,r_{m2}}~. \end{eqnarray} If the condition 

\begin{eqnarray}\label{condition} |r_{m2} \,-\, r_{m1}| \, \ll \, \left|k(\bar{\omega}\,+\,\delta \omega) \,-\, k(\bar{\omega}) \right|^{-1} ~,\end{eqnarray}

\noindent is satisfied, then one can write

\begin{eqnarray}  A_m(t,\,r_{m1})A^*_m(t,\,r_{m2})
 \,\approx \, A_m(t,\,r_{m1})A^*_m(t,\,r_{m1}) \,\approx \nonumber \\  A_m(t,\,r_{m2})A^*_m(t,\,r_{m2})~.~~~ \end{eqnarray} Namely, using Eq.\eqref{amplitude} one observes that ($\delta r \equiv r_2 -r_1 $) 

\begin{widetext}
\begin{eqnarray} &&  A_m(t,\,r_{m1})A^*_m(t,\,r_{m2}) \,=\,  \int\limits_{\bar{\omega}-\delta\omega /2 }\limits^{
\bar{\omega}+\delta\omega / 2}d\omega d\omega' \,
a(\omega)a^*(\omega') \,e^{-i \left(\omega - \omega'\right)
    t } \,e^{i \left[k(\omega) -
      k(\omega')\right] \,r_{m1} } e^{-i \left[k(\omega') -
      k(\bar{\omega})\right] \,\delta r }  \,\approx \nonumber \\&&  \int\limits_{\bar{\omega}-\delta\omega /2 }\limits^{
\bar{\omega}+\delta\omega / 2}d\omega d\omega' \,
a(\omega)a^*(\omega') \,e^{-i \left(\omega - \omega'\right)
    t } \,e^{i \left[k(\omega) -
      k(\omega')\right] \,r_{m1} }  \,=\, A_m(t,\,r_{m1})A^*_m(t,\,r_{m1}) ~, \nonumber  \end{eqnarray}\end{widetext} and analogously one finds that this expression is pretty much the same as $A_m(t,\,r_{m2})A^*_m(t,\,r_{m2})$. It is worth noticing that if the Eq.\eqref{condition} is satisfied in the case of standard dispersion relation, then it is automatically satisfied for the dispersion relations Eqs.(\ref{moddisprel}, \ref{meoredisperstanafardoba}) for $dk/d\omega <1$ in both cases when $\beta > 0$. Thus, assuming the condition \eqref{condition} is satisfied, we may write
\begin{eqnarray}\label{corfunc}  &&\langle\varPhi_1(t)\varPhi^*_2(t) \rangle \,= \nonumber \\ && \sum\limits_m \left\langle  A_m(t,\,r_{m1}) A^*_m(t,\, r_{m1})  \right\rangle
  \,{e^{ik(\bar{\omega})(r_{m1}-r_{m2})}\over
  r_{m1}\,r_{m2}}~.~~~~ \end{eqnarray} The quantity $\langle A_m(t,\,r_{m1}) A^*_m(t,\, r_{m1}) \rangle$ characterizes the intensity of the radiation
from the
  source element $d\sigma_m$. So the correlation function Eq.(\ref{corfunc}) takes
the form \begin{equation}\label{corfunc1} \langle\varPhi_1(t)\varPhi^*_2(t) \rangle \,=\,
  \int\limits_{\Sigma} d\sigma I(\sigma)
\,\frac{e^{ik(\bar{\omega})(r_{1}-r_{2})} }{
  r_{1}\,r_{2}}~, \end{equation} where $I(s_m)d\sigma_m =
\langle
  A_m(t,\,r_{m1}) A^*_m(t,\, r_{m1})\rangle $. In most applications the intensity
$I(\sigma)$ may be
  assumed to be uniform on $\Sigma$. To work out the integral Eq.(\ref{corfunc1}) let us
denote by
  $(\xi,~\eta)$ the coordinates of a point $S$, referred to axes at
$O$, and
  let $(x_1,~y_1)$ and $(x_2,~y_2)$ be the coordinates of $P_1$ and
$P_2$
  referred to parallel axes at $O'$, see Fig.1. Retaining only leading terms in
$x/r,~y/r,~\xi/r,~\eta/r$, where $r$ is the distance $OO'$, one finds
  \begin{eqnarray} && r_1 \,-\, r_2  \,\approx \nonumber \\&&  \frac{ x_1^2 \,+\, y_1^2 \,-\, x_2^2 \,-\, y_2^2 \,+\, 2\,\xi \, (x_2 \,-\,
    x_1) \,+\, 2 \, \eta \, (y_2 - y_1) }{ 2\,r} ~. \nonumber \end{eqnarray}

\noindent   Denoting \begin{eqnarray}&& p \,=\, \frac{x_1 \,-\, x_2
}{ r
  }~,~~~~q \,=\, \frac{y_1 \,-\, y_2 }{ r}~,  \nonumber \\ && \psi \,=\, \frac{k(\bar{\omega}) (x_1^2 \,+\, y_1^2 \,-\,
x_2^2 \,-\,
    y_2^2) }{ 2r }~, \label{phasepsi} \end{eqnarray} the Eq.(\ref{corfunc1}) takes the form
\begin{equation}\label{corfunc2}\langle\varPhi_1(t)\varPhi^*_2(t) \rangle \,\approx \,
\frac{e^{i\psi}}{ r^2}
  \int\limits_{\Sigma} d\xi d\eta I(\xi,~\eta) \,e^{ik(\bar{\omega})(p\xi
-
    q\eta)}~. \end{equation} For a uniform circular source of
radius $\rho$
with its center at $O$, the Eq.(\ref{corfunc2}) reduces to

  \begin{equation}\label{mutcoh} \langle\varPhi_1(t)\varPhi^*_2(t)
\rangle \,\sim \,  e^{i\psi}  \,\frac{J_1(v)}{v}~, \end{equation} where $v=k(\bar{\omega})\rho\sqrt{p^2 + q^2} $ and
$J_1$ stands for the first kind and first order Bessel function. In most
applications
  the quantity $\psi$ is very small, so that one can neglect
corresponding phase factor
  in Eq.(\ref{mutcoh}). The function
  $J_1(v)/v$ decreases steadily from the value $0.5$ when $v=0$ to the
value
  zero when $v=3.83$ indicating that the degree of coherence steadily
  decreases and approaches complete incoherence when $P_1$ and $P_2$
are
  separated by the distance \[P_1P_2  \,=\, \frac{0.61\, \bar{\lambda}\, r }{
    \rho}~,\] where $\bar{\lambda} = 2\pi / k(\bar{\omega})$. In experiments on interference and diffraction a
departure of
  $12$ per cent from the ideal value of coherence that occurs at $v=1$
can be
  taken as a maximum permissible departure that gives for the
separation of
  points $ P_1$ and $P_2$ \begin{equation}\label{explim} P_1P_2 \,=\,
\frac{0.16\, \bar{\lambda}\, r }{\rho}~.\end{equation}

\section{Discussion }

Present paper is a continuation of the discussion started in \cite{Maziashvili:2009gt}. We have constructed the Hilbert space representation for a relatively broad class of minimum-length deformed position-momentum uncertainty relation \eqref{gur}, which in its turn can be understood as a modification arising because of quantum-gravitational fluctuations of the background Minkowski space. A particular case of this sort of $\mathsf{PL}\mathcal{QM}$ is a well known example studied in \cite{Kempf:1996nk, Kempf:1996fz}. Following this construction, we have then used $\mathsf{PL}\mathcal{QM}$ for constructing the correspondingly modified field theory, which can be used for estimating corrections to various quantities. We have applied the minimum-length deformed field theory constructed this way for estimating the corrections to the light coherence rate from distant astrophysical objects. The rate of light (in)coherence is estimated on the basis of van Cittert-Zernike formalism (as pointed out in \cite{Coule:2003td}). The picture arising from this consideration looks as follows.

If all the assumptions required for light coherence is satisfied in the case of standard dispersion relation, then those conditions are automatically satisfied for modified dispersion relations (\ref{moddisprel}, \ref{meoredisperstanafardoba}). Namely, if the requirement - the path difference $|r_{m2} - r_{m1}|$ to be smaller than the duration of the wave packet $\delta t \equiv \delta \omega^{-1}$ is satisfied in the standard case, that is, $|r_{m2} - r_{m1}| \ll \delta\omega^{-1}$, then Eq.\eqref{condition} is automatically satisfied for dispersion relations (\ref{moddisprel}, \ref{meoredisperstanafardoba}) as $dk/d\omega < 1$ in those cases. 

In the standard case the phase $\psi$ in Eq.\eqref{mutcoh} is usually neglected as it is usually small in most applications. In the cases of Eq.(\ref{moddisprel}, \ref{meoredisperstanafardoba}) $\psi$ becomes even smaller as $k(\bar{\omega}) < \bar{\omega}$.

And finally, the maximum separation of points $P_1$ and $P_2$ over which light is coherent, Eq.\eqref{explim}, becomes enlarged as compared to the standard case for the wave-length $\bar{\lambda}$ is enlarged now: $1/k(\bar{\omega}) > 1/\bar{\omega}$.

One can simply estimate the rate of the effect in both cases: $\alpha=2$ and $\alpha=1/2$. Restricting ourselves to the leading order corrections, in the former case ($\alpha = 2$) one finds from Eq.\eqref{moddisprel} (as well as from the relation $\delta \lambda = \beta l_P^{\alpha}\lambda^{1-\alpha}$ by which we started our discussion, see the Introduction) that the wave-length increment takes the form $\delta \bar{\lambda} \simeq  l_P^2 \bar{\omega}$. Let us see if the increment in separation $P_1P_2$ can be made observationally perceptible, say of the order of $1$\,cm. Even if we take $\bar{\omega} \simeq E_P$, from Eq.\eqref{explim} one gets $r /\rho \sim 10^{33}$. To obtain such a huge ratio is absolutely impossible for the present horizon radius is about $10^{28}$\,cm and in its turn $\rho$ is by many orders of magnitude greater than $1$\,cm. So for the realistic values of $\bar{\omega},\, r$ and $\rho$ one gets a minuscule effect. 

Analogously, in the latter case $\alpha=1/2$ one finds from Eq.\eqref{meoredisperstanafardoba} (as well as from the relation $\delta \lambda = \beta l_P^{\alpha}\lambda^{1-\alpha}$ by which we started our discussion, see the Introduction) that $\delta \bar{\lambda} \simeq \left(l_P \bar{\lambda}\right)^{1/2}$. Let us again consider an extreme case to demonstrate the smallness of the effect. Let us take $P_1P_2 \simeq 10^6$\,cm and assume that we can measure this distance with accuracy $\sim 1$\,cm. Taking $\bar{\lambda} \simeq 10^{-9}$\,cm, from Eq.\eqref{explim} one finds \[\delta \left(P_1P_2\right) \,\simeq \, P_1P_2 \, \left( \frac{l_P}{\bar{\lambda}} \right)^{1/2} \, \simeq \, 10^{-6}\text{cm}~. \] So, in any realistic case we expect the effect to be tiny.

For other approaches describing the effect of quantum-gravity on the light coherence from distant astrophysical objects (indicating that the effect is intangible) see \cite{Chen:2006qj, Dowker:2010pf, Tamburini:2011yf}.


\begin{thebibliography}{10}


\bibitem{LL}

L.~D.~Landau and E.~M.~Lifshitz, "The Classical Theory of Fields" (Landau and Lifshitz Course of Theoretical Physics - Vol. II; Butterworth-Heinemann, Oxford, 1987). 
         

\bibitem{Osborne:1949zz}
  M.~F.~M.~Osborne,
  Phys.\ Rev.\  {\bf 75} (1949) 1579.


\bibitem{Anderson}

J.~L.~Anderson, Rev.\ Mex.\ Fis.\ {\bf 03(3)} (1954) 176. 



\bibitem{Wigner:1957ep}
  E.~P.~Wigner,
  Rev.\ Mod.\ Phys.\  {\bf 29} (1957) 255.
  
  
\bibitem{Salecker:1957be}
  H.~Salecker and E.~P.~Wigner,
  Phys.\ Rev.\  {\bf 109}, 571 (1958).


\bibitem{Regge:1958wr}
  T.~Regge,
  Nuovo Cim.\  {\bf 7}, 215 (1958).
  

\bibitem{Peres:1960zz}
  A.~Peres and N.~Rosen,
  Phys.\ Rev.\  {\bf 118}, 335 (1960).
  
 
\bibitem{Mead:1964zz}
  C.~A.~Mead,
  Phys.\ Rev.\  {\bf 135}, B849 (1964).
  
  
\bibitem{Mead:1966zz}
  C.~A.~Mead,
  Phys.\ Rev.\  {\bf 143}, 990 (1966).
  
  


\bibitem{Karolyhazy:1966zz}
  F.~Karolyhazy,
  Nuovo Cim.\  A {\bf 42} (1966) 390.



\bibitem{Planck}
M.~Planck, 
Sitzungsberichte der K\"{o}niglich Preußischen Akademie der Wissenschaften zu Berlin, 1899 (I) Page 479. 
  


\bibitem{Garay:1994en}
  L.~J.~Garay,
  Int.\ J.\ Mod.\ Phys.\  A {\bf 10}, 145 (1995)
  [arXiv: gr-qc/9403008].





\bibitem{Lieu:2003ee}
  R.~Lieu and L.~W.~Hillman,
  Astrophys.\ J.\  {\bf 585}, L77 (2003)
  [arXiv: astro-ph/0301184].


\bibitem{Ng:2004xr}
  Y.~J.~Ng,
  Lect.\ Notes Phys.\  {\bf 669}, 321 (2005)
  [arXiv: gr-qc/0405078].


\bibitem{Maslowski:2012aj} 
  T.~Maslowski, A.~Nowicki and V.~M.~Tkachuk,
  J.\ Phys.\ A A {\bf 45}, 075309 (2012)
  [arXiv: 1201.5545 [quant-ph]].


\bibitem{Kempf:1996nk}
  A.~Kempf and G.~Mangano,
  Phys.\ Rev.\  D {\bf 55}, 7909 (1997)
  [arXiv: hep-th/9612084].

\bibitem{Kempf:1996fz}
  A.~Kempf,
  J.\ Phys.\ A  {\bf 30}, 2093 (1997)
  [arXiv: hep-th/9604045].


 




 



\bibitem{BornWolf}
P.~H.~van~Cittert, Physica\ {\bf 1} (1934) 201; 

P.~H.~van~Cittert, Physica\ {\bf 6} (1939) 1129;

F.~Zernike, Physica\ {\bf 5} (1938) 785; 

L. Janossy, Nuovo\ Cimento\ {\bf 6}
(1957) 111; 

L. Janossy, Nuovo\ Cimento\ {\bf 12} (1959) 369; 

H. H. Hopkins, Proc.\ Roy.\
Soc.\ (London)\ A {\bf 208} (1951) 263; 

H. H. Hopkins, Proc.\ Roy.\ Soc.\ (London)\ A {\bf 217}
(1953) 408; 

M. Born and E. Wolf, "Principles of Optics",
(Cambridge University Press, 2002). 


\bibitem{Maziashvili:2009gt}
  M.~Maziashvili,
  Astropart.\ Phys.\  {\bf 31}, 344 (2009)
  [arXiv: 0901.2405 [gr-qc]].

 
 
\bibitem{Coule:2003td}
  D.~H.~Coule,
  Class.\ Quant.\ Grav.\  {\bf 20}, 3107 (2003)
  [arXiv: astro-ph/0302333].



\bibitem{Chen:2006qj}
  Y.~Chen and L.~Wen,
  arXiv: gr-qc/0605093.

\bibitem{Dowker:2010pf}
  F.~Dowker, J.~Henson and R.~Sorkin,
  Phys.\ Rev.\  D {\bf 82}, 104048 (2010)
  [arXiv: 1009.3058 [gr-qc]].



\bibitem{Tamburini:2011yf}
  F.~Tamburini, C.~Cuofano, M.~Della Valle and R.~Gilmozzi,
  Astron.\ Astrophys.\  {\bf 533}, A71 (2011)
  [arXiv: 1108.6005 [gr-qc]].
  
 


  
  


\end{thebibliography}
\end{document}